# Dual-Interrogation Method for Suppressing Light Shift in Rb 778 nm Two-Photon Transition Optical Frequency Standard

Dou Li[1,2], Kangqi Liu[1,2], Pengfei Wang[2], and Songbai Kang[2*]

[1] *School of Physical Sciences, University of Chinese Academy of Sciences, Beijing 100049, China.*

[2] *Key Laboratory of Atomic Frequency Standard,*

*Innovation Academy for Precision Measurement Science and Technology, Chinese Academy of Sciences, Wuhan 430071, China*

In this study, a dual-interrogation (DI) method was used to suppress the light shift in the Rb 778 nm $5S_{1/2} \rightarrow 5D_{5/2}$ two-photon transition (TPT) optical frequency standard. The approach used an auxiliary system to calibrate the light shift of the primary system in real time to mitigate the absolute light shift and suppress the sensitivity of the system to the optical power. Results show that after using the DI method, the absolute light shift and light-power sensitivity of the system were reduced by a factor of 10. The proposed method will improve the accuracy of the Rb 778 nm TPT optical frequency standard and increase the mid- and long-term stability. The method can also be applied to other vapor-cell atomic frequency standards that experience light shifts.

## I. INTRODUCTION

Continuous improvements in optical atomic clock performance and the growing need for frequency metrology and highly accurate timekeeping have propelled a recent surge in research on compact optical atomic clocks [1]. The Rb 778 nm $5S_{1/2} \rightarrow 5D_{5/2}$ two-photon transition (TPT) has attracted particular attention as an optical frequency standard [2–3]. Two-photon spectroscopy technology has no atomic velocity selection and reduces the first-order Doppler shift and broadening [4]. Therefore, a Rb 778 nm TPT frequency standard based on vapor cell technology has achieved an excellent signal-to-noise ratio with a mm-sized vapor cell and without complex atomic cooling procedures [5–10]. Moreover, unlike the two-color Rb $5S_{1/2} \rightarrow 5D_{5/2}$ TPT scheme [11], the Rb 778 nm TPT requires only a single laser source, resulting in a simpler optical setup and facilitating miniaturization. The 778 nm wavelength can be accessed through reliable frequency doubling of 1556 nm lasers in the stable communication C-band. Consequently, the Rb 778 nm TPT has been endorsed by the BIPM [3] for definition of the meter through optical transitions and is viewed as a potential optical transition for future portable and next-generation chip-scale optical clocks. Currently, high-performance Rb 778 nm two-photon transition optical frequency standards (TPTOFS) with stabilities of $3\times10^{-13}\tau^{-1/2}$ (1–10000 s) and $1.8\times10^{-13}\tau^{-1/2}$ (<100 s) have been reported [4,6]. Regarding portable applications, a Rb 778 nm optical clock whose critical components are integrated devices has been demonstrated and showed a stability of $4.4\times10^{-12}\tau^{-1/2}$ [9]. A miniaturized Rb TPTOFS implemented on a micro-optics breadboard with a volume of approximately 35 ml$^3$ has also been reported with a stability performance of $2.9\times10^{-12}\tau^{-1/2}$ (<1000 s) [7]. Among these TPTOFSes, light shift is one of the primary limiting factors for long-term stability and accuracy. While methods for suppressing light shifts in trapping systems and atomic pumping have been widely explored (e.g., magic wavelengths [12] and quadrature locking methods [13]), strategies for eliminating the shift caused by interrogating light have seldom been reported. A typical method is to stabilize the light power to weaken the light shift effect on the long-term performance of the TPTOFS [14]. However, a more effective solution would be to decrease the sensitivity of the clock transition to light power, that is, the light-shift coefficient.

---

* kangsongbai@apm.ac.cn

Yudin et al. proposed a technique for mitigating atomic clock light shifts using a power modulation (PM) combined error signal (CES) [15]. This method sequentially interrogates atoms with two distinct optical powers in a single modulation cycle. This results in the generation of a CES with a zero-point amplitude corresponding to the zero-light-shift frequency of the atomic clock, thereby ensuring frequency locking free of light shifts. However, this approach overlooked the impact of laser frequency noise within a single power modulation cycle. Given that the laser frequency operates freely within the modulation cycle, any fluctuations in the atom interrogation frequency during this cycle can result in a CES that does not accurately represent the zero-light-shift frequency of the atomic clock. In extreme cases, this can lead to locking failure. To date, this method has not been successfully implemented in TPTOFS experiments.

To address the light-shift problem, this paper presents a dual interrogation (DI) method. This method involves the simultaneous interrogation of the primary 778 nm TPT atomic system and an auxiliary atomic system, with a single laser [16]. This results in a real-time combination of the two error signals, forming a CES. This CES is then used for continuous feedback and control of the laser frequency, suppressing the light shifts of the primary atomic system. The critical difference between the proposed DI approach and the PM-CES method is the use of spatial instead of temporal interrogation to achieve real-time regulation of the laser frequency. This technique avoids the effect of laser frequency noise, thereby making the frequency loop-locking process more robust. In principle, this approach can suppress the sensitivity of the 778 nm TPTOFS to fluctuations in light power and reduce the influence of light shifts on its frequency accuracy.

## II. PRINCIPLE OF THE DI METHOD

In Fig. 1, the horizontal axis represents the laser oscillator frequency $f_L$ and the vertical axis represents the amplitude of the error signals. $S_1(P_1, f_L)$, $S_2(P_2, f_L)$, and $S_{ces}(P_1, P_2, f_L)$ correspond to the error signals of the primary atomic system (#1), auxiliary system (#2), and CES, respectively; $f_{01}, f_{02}, f_1, f_2, \delta f_{P1}, \delta f_{P2}, \tan\varepsilon_1$, and $\tan\varepsilon_2$ correspond to the zero-light-shift frequency, output locking frequency, light shifts, and slope of error signal, respectively, for systems #1 and #2 operating independently; $\Delta v$ is the residual difference between the zero-light-shift frequency of the two systems, defined as $\Delta v \equiv f_{01} - f_{02}$, a discrepancy which may arise from inconsistencies between the two atomic systems, such as the residual gas pressure of the Rb vapor cell and the absolute temperature differences; and $V_{01}$ and $V_{02}$ are the amplitudes of the respective error signals when the local oscillator frequency is $f_{01}$, i.e., $V_{01} = S_1(P_1, f_{01})$ and $V_{02} = S_2(P_2, f_{01})$.

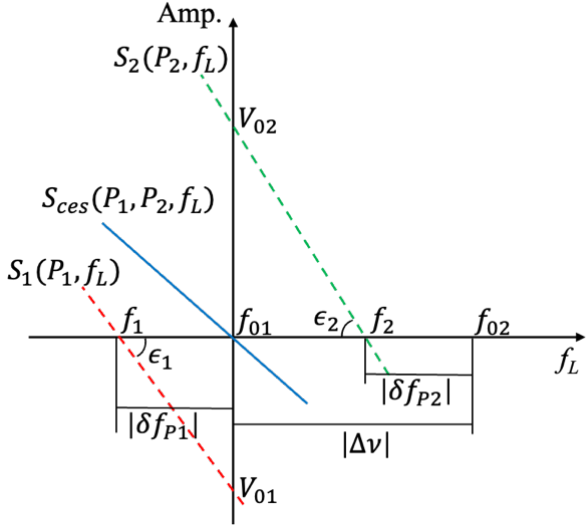

FIG. 1. Schematic of error signal linear region. Red dashed line is the error signal for the primary #1 system and the green dashed line is the error signal for the auxiliary #2 system, the blue solid line is the CES.

To make the locking frequency of $S_{ces}$ ($P_1$, $P_2$, $f_L$) equal to $f_{01}$, the following equation must be satisfied:

$$S_{ces}(P_1, P_2, f_{01}) = S_1(P_1, f_{01}) + \beta S_2(P_2, f_{01}) = 0, \quad (1)$$

that is

$$V_{01} + \beta V_{02} = 0, \quad (2)$$

where $\beta$ is the combination factor. To calculate $\beta$, we need to make three assumptions for the Rb TPT, similar to the one in Ref. [15]:

1. The light shift is linearly related to the interrogation power. That is,

$$\delta f_{Pi} = C_i P_i. \quad (3)$$

2. The slope of the TPT signal is consistently proportional to the squared power, meaning there is no power broadening [17]. That is,

$$\tan \varepsilon_i = b_i P_i^2. \quad (4)$$

3. The light shifts for systems #1 and #2 are significantly smaller than the linewidth of the spectral line. That is,

$$\delta f_{Pi} \ll \gamma. \quad (5)$$

In Eq. (3)–(5), $i$ represents the $i$-th system ($i$=1, 2), $C_i$ and $b_i$ are the light shift coefficient and the characteristic parameter for system $i$, respectively. The parameter $b_i$ is associated with factors such as the diameter of the interrogation beam, atomic number density, fluorescence collection efficiency, and quantum efficiency of the detector. $\gamma$ represents the linewidth of the Rb TPT.

Based on the geometric relationship depicted in Fig. 1, we can derive

$$V_{0i} = (f_{01} - f_i) \tan \varepsilon_i \quad (i = 1,2). \quad (6)$$

Substituting Eq. (2), (4), and (5) into Eq. (1) gives

$$b_1 P_1^2 C_1 P_1 + \beta(f_{02} + C_2 P_2 - f_{01}) b_2 P_2^2 = 0, (7)$$

and

$$\beta = -\frac{V_{01}}{V_{02}}$$

$$= \frac{b_1 C_1 P_1^3}{(\Delta \nu - C_2 P_2) b_2 P_2^2}, (\Delta \nu - C_2 P_2 \neq 0). \quad (8)$$

Finally, the CES can be expressed as

$$S_{ces}(P_1, P_2, f_L) = S_1(P_1, f_L) + \beta S_2(P_2, f_L). \quad (9)$$

It is important to clarify that $S_2(P_2, f_L)$ was exclusively employed for the absolute frequency calibration of the primary system (#1). This did not affect the error signal slope of system #1. The short-term frequency stability limit of the system using the DI method was determined by the performance parameters of primary system #1 ($P_1$ and $b_1$).

### III. EXPERIMENTAL SETUP

The experimental setup is illustrated in Fig. 2. A 1556 nm fiber laser output was split into two beams. The first beam was used for beating with an optical frequency comb, referenced by a hydrogen maser (for absolute frequency measurements) or an ultrastable cavity (for stability measurements), and the second was amplified by an erbium-doped fiber amplifier (EDFA) and processed by a second-harmonic generation (SHG) unit before being used to interrogate the atom. The second beam was further split in two, each with a power of approximately 15 mW, for interrogation in systems #1 and #2. The power of each beam was adjusted using variable optical attenuators (VOA1 and VOA2). Glan–Taylor prisms (GTPs) ensured that the probing light maintained a high-purity linearly polarized state. Photodetectors PD1 and PD2 monitored the power of the interrogated light.

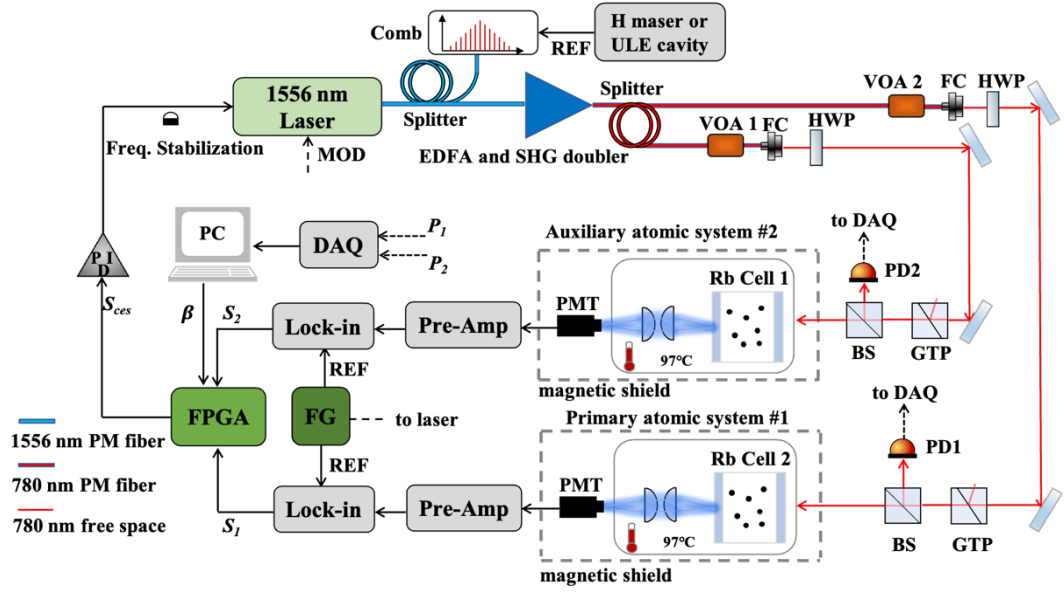

FIG. 2. Experimental setup of the DI method. Terms used: VOA, variable optical attenuator; FC, fiber collimator; HWP, half-wave plate; PD, photodiode; PID, proportional-integral-derivative; BS, beam splitter; GTP, Glan–Taylor prism; PMT, photomultiplier tube; Lock-in, lock-in amplifier; Pre-Amp, preamplifier; FPGA, field-programmable gate array; EDFA, erbium-doped fiber amplifier; SHG, second harmonic generation; FG, function generator; and DAQ, data acquisition.

In the two atomic systems, cubic rubidium vapor cells with a dimension of approximately 6 mm were temperature-controlled at around 97 °C. The rear surfaces of the rubidium cells were coated with 778 nm high-reflectivity films, enabling the efficient reflection of incident photons to guarantee the TPT's special mode matching. A pair of aspherical lenses within the physical packages and a photomultiplier tube (PMT) were used to collect and detect the 420 nm fluorescence signal. Magnetic shielding effectively isolated the vapor cells from environmental magnetic fields, preventing undesired Zeeman splitting of atomic energy levels.

To lock the laser frequency, a modulation frequency of 20 kHz with a modulation depth of approximately 50 kHz, was applied to the laser source. The output of the atomic signal was amplified using a preamplifier. A lock-in amplifier demodulated the frequency-modulated atomic signal and outputted the DC error signal. The two separated beams were utilized to interrogate systems #1 and #2, generating error signals denoted as $S_1$ and $S_2$. Power monitor signals from PD1 and PD2, obtained via a data acquisition, were employed to compute $\beta$. Finally, $S_1$, $S_2$, and $\beta$ were passed to a homemade field-programmable gate array to generate $S_{ces}$, which was subsequently provided to a proportional-integral-derivative controller to accomplish closed-loop frequency locking.

## IV. RESULTS

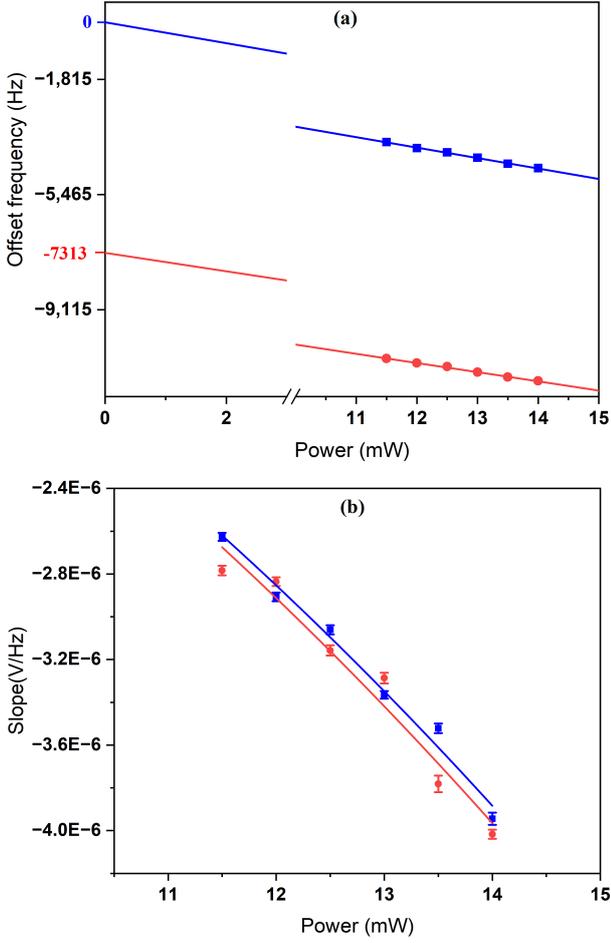

FIG. 3. (a) Light shift characteristics of the 778 nm interrogation light for systems #1 (solid blue rectangles) and #2 (solid red circles). The solid lines represent a linear fit, and the discontinuities are due to the use of break processing on the x-axis. The zero-point offset frequency is the zero-light-shift frequency of system #1, $f_{01}$. (b) Error signal slope as a function of light power for systems #1 and #2. The solid lines are the fitted curves of Eq. (4).

During the experiment, we set $\beta=0$ to measure the parameters (light shift, zero-light-shift frequency, slope of the error signal, etc.) of system #1. For system #2 measurements, we set $\beta=1$ and shielded against the interrogating light of system #1. The electronic gain of the locking loop should be carefully optimized to avoid electronic frequency shifts during measurement. The dependence of the output frequency and slope amplitude of the error signal on the interrogation light power for systems #1 and #2 are shown in Fig. 3(a) and 3(b), respectively. $C_1$, $C_2$, $f_{01}$, $f_{02}$, their fitting errors obtained by linear fitting, and $b_1$, $b_2$, and their fitting errors obtained by quadratic fitting are summarized in Table I. The frequency difference $\Delta v$ between $f_{01}$ and $f_{02}$ is 7312±149 Hz, resulting mainly primarily from the inconsistent residual background gas pressures within the atomic cells [18].

Table I. Measured physical parameters of systems #1 and #2.

| System | $C_i$ (Hz/mW) | $b_i$ (V·Hz$^{-1}$·mW$^{-2}$) | $f_{0i}$ (Hz) |
|---|---|---|---|
| #1 | −331 (9) | −0.02 (1.3E-4) | 0 |
| #2 | −291 (8) | −0.02 (2.4E-4) | -7312 (149) |

Note: Numbers in parentheses are fitting errors.

To obtain the light-shift characteristic of system #1, operated under a DI mode, we fixed the light power of system #2 to 14 mW and increased the power of system #1 from 11.5 mW to 14 mW in 0.5 mW steps using a VOA 1. From the physical parameters outlined in Table I, we calculated the corresponding $\beta$ value for each power combination and generated the CES to stabilize the laser output frequency. The absolute locking frequency was measured using an optical frequency comb referenced by a hydrogen maser. Figure 4 shows the light-shift characteristics for system #1 with (orange) and without (blue) applying the DI method. The zero-offset frequency on the vertical axis represents the fitted zero-light-shift frequency of system #1 ($f_{01}$). The DI method data did not exhibit linear characteristics, indicating significant suppression of the light shift.

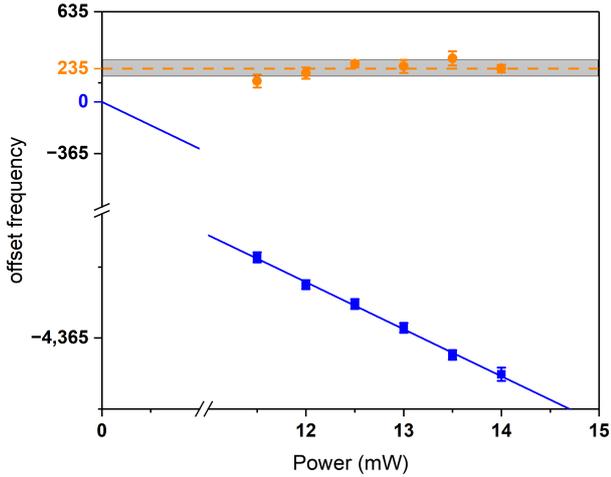

FIG. 4. Comparison of the light shift characteristics of system #1 with (solid orange circles) and without (solid blue rectangles) the DI method. The solid lines represent a linear fit, and the discontinuities are due to the use of break processing on both the x- and y-axes. The dashed line represents the average frequency from the DI method with the standard deviation represented by the gray shading. The zero-point offset frequency is the zero-light-shift frequency of system #1, $f_{01}$.

The light-shift characteristic data from the DI method have an average value of 235 Hz with a standard deviation of 55 Hz. The reason for the discrepancy between the average value of the data and the zero-offset frequency (or $f_{01}$) (see Table I) was mainly from the fitting error of $f_{01}$ and the uncertainty of the zero-power point (3% precision of Thorlabs's power meter), with a small contribution from the error in $\beta$. However, the deviation indicates the repeatability of the system when running in the DI mode. Because we continuously turned the loop on and off during a one-day measurement, the DI scheme may contribute to repeatability owing to the voltage changes in the CES electronic unit. In addition, the repeatability of the atomic system may be in part owed to environmental factors, such as temperature and magnetic field fluctuations. Further in-depth investigation into the repeatability of the DI method is required in the future.

To further demonstrate the real-time mitigation effect of the DI scheme on the light shift, we added an interruption to the interrogation power of system #1 to simulate power fluctuation and measured the changes in the locking frequency of the system before and after using the DI method. We actively adjusted the interrogating power of system #1 using VOA1 and maintained the interrogating power of system #2 at 14 mW. The power underwent a slow change (triangular wave) with an amplitude ranging from 9–16 mW for 1000 s, as shown in Fig. 5(a). Figure 5(b) illustrates the variation in the system output frequency under the corresponding power-interruption period, both with and without the DI scheme. In the absence of DI, the output frequency fluctuated noticeably in line with the interrupted light power, with an average offset frequency of approximately −4865 Hz. Conversely, with DI, the output frequency remained constant and was unaffected by light power variations. Furthermore, the central offset frequency with the DI method was approximately 411 Hz, successfully reducing the light shift by one order of magnitude to the sub-kilohertz level. The difference between the central frequency and the result shown in Fig. 4 is attributed to the repeatability of the system, as this real-time demonstration was performed 10 days after the previous characteristic measurement. Figure 5(c) shows the corresponding Allan deviation results for the output frequency with (black) and without (red) applying the DI method. The fiber comb was referenced by an ultrastable laser to accurately evaluate the frequency stability performance (green triangles in Fig. 5). The stability results show that, without the DI method, the light shift becomes the main contribution to instability, seriously degrading the stability of the system. However, when the DI method was used, the light-shift effect was effectively suppressed. At the 100-second averaging time, the system stability differs by one order of magnitude between using and not using the DI scheme. Under conditions of the same light power interference, this result implies that the light-shift coefficient of the system is reduced by a factor of 10 when using the DI scheme. For averaging times of 1 s, the stability with and without the DI method is consistent, demonstrating that the DI method does not affect the short-term performance of the system.

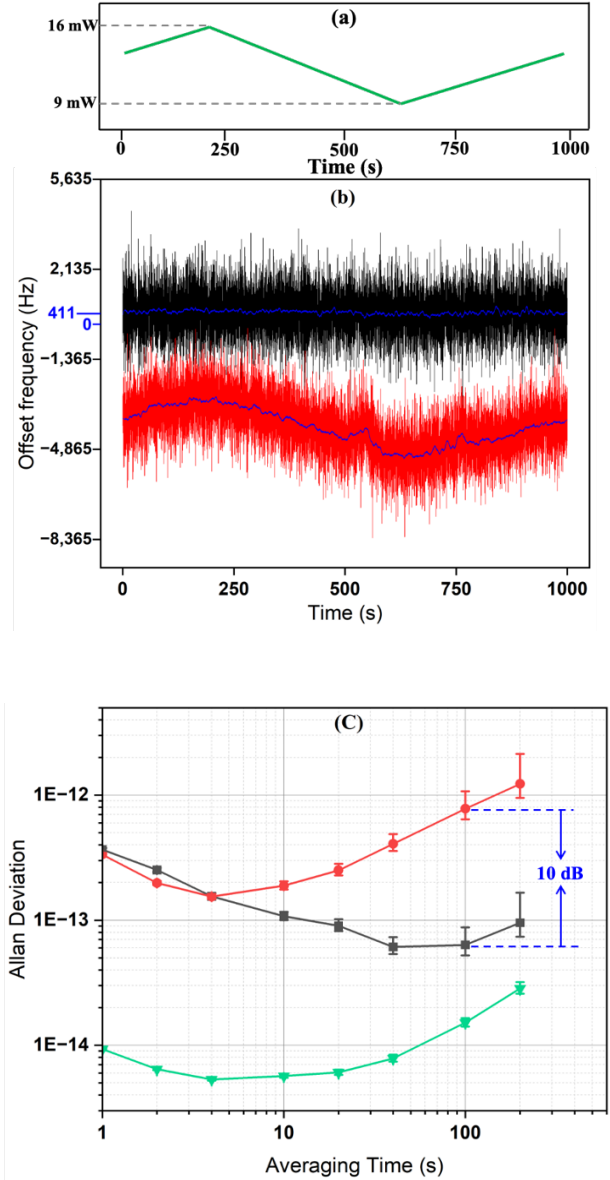

FIG. 5. (a) Laser power variation, (b) output frequency, and (c) Allan deviation of system #1 under a triangular wave perturbation with a power variation range of 9–16 mW and a period of 1000 s. Black represents data obtained with the DI method and red represents the original data of system #1. The blue lines in (b) are 100-points smoothed data from the original data. The green triangles represent the performance of the optical reference. The zero-point offset frequency is the zero-light-shift frequency of system #1, $f_{01}$.

## V. CONCLUSIONS AND DISCUSSION

We proposed and experimentally demonstrated a dual-interrogation (DI) method for suppressing the light shift in a Rb 778 nm TPTOFS. When the primary system was operated with an auxiliary system using the DI method, the absolute light shift and light power sensitivity of the system were significantly suppressed by more than one order of magnitude compared with an independently operated system. This indicates that the DI method can enhance the accuracy of the TPTOFS and its mid- and long-term stabilities. For future practical applications, the repeatability of the DI method deserves further investigation, especially regarding the contribution from electronics for combining error signals and the PD gain for power monitoring. Considering that a chip-scale atomic system has been demonstrated [7,10], the utility of the second auxiliary system will not significantly impact the complexity, volume, or power consumption of the TPTOFS system. In principle, the DI method can also be used for other vapor-cell standards that experience light shifts, such as one-photon sub-Doppler spectroscopy and Doppler-free spectroscopy of molecular iodine [19,20].

## ACKNOWLEDGEMENT

We thank Qunfeng Chen for providing the ultrastable cavity stabilized laser and John

Kitching for suggesting DI method's dynamic demonstration.